\documentclass[aps,preprint,superscriptaddress,]{revtex4}

\usepackage{amsfonts}
\usepackage{amsmath}
\usepackage{amssymb}
\usepackage{graphicx}
\usepackage{mathrsfs}
\begin{document}

\title{Unconditional security proof of a deterministic quantum key distribution with
a two-way quantum channel}
\author{Hua Lu}
\affiliation{State Key Laboratory of Magnetics Resonances and Atomic
and Molecular Physics, Wuhan Institute of Physics and Mathematics,
Chinese Academy of Sciences, Wuhan 430071, People's Republic of
China}
\affiliation{Department of Mathematics and Physics, Hubei
University of Technology, Wuhan 430068, People's Republic of China}
\author{Chi-Hang Fred Fung}
\affiliation{Department of Physics and Center of Computational and Theoretical
 Physics, University of Hong Kong, Pokfulam Road, Hong Kong}
\author{Xiongfeng Ma}
\affiliation{%
Center for Quantum Information and Quantum Control,\\
Department of Physics, University of Toronto, Toronto, Canada, M5S 1A7 \\
}
\author{Qing-yu Cai}
\email{qycai@wipm.ac.cn}
\affiliation{State Key Laboratory of
Magnetics Resonances and Atomic and Molecular Physics, Wuhan
Institute of Physics and Mathematics, Chinese Academy of Sciences,
Wuhan 430071, People's Republic of China}


\begin{abstract}

In a deterministic quantum key distribution (DQKD) protocol with a
two-way quantum channel, Bob sends a qubit to Alice who then encodes
a key bit onto the qubit and sends it back to Bob. After measuring
the returned qubit, Bob can obtain Alice's key bit immediately,
without basis reconciliation. Since an eavesdropper may attack the
qubits traveling on either the Bob-Alice channel or the Alice-Bob
channel, the security analysis of DQKD with a two-way quantum
channel is complicated and its unconditional security has been
controversial. This paper presents a security proof of a
single-photon four-state DQKD against general attacks.

\end{abstract}

\volumeyear{year}
\volumenumber{number}
\issuenumber{number}
\eid{identifier}
\date{\today}
\maketitle

\section{Introduction}
Quantum cryptography enables two remote parties to share an
information-theoretically secure key, which can be used for later
cryptographic applications. Since the pioneering protocol was
presented by Bennett and Brassard in 1984 (called the BB84 protocol)
\cite{bb84}, its security against general attacks has been studied
in idealized settings~\cite{mayer01, lochau99, sp00} and also in
practical settings~\cite{sbcd}. Meanwhile, a deterministic quantum
key distribution (DQKD) protocol with a two-way quantum channel has
been proposed \cite{bf02,cai03}, which allows Alice and Bob to
encode and decode secret message in a deterministic manner. Bob can
obtain Alice's key bit directly with his measurement outcomes,
without a basis reconciliation step, which makes key distribution
efficient and even a quasi-secure direct communication possible when
the two parties are connected with an ideal two-way quantum channel
\cite{bf02,cai04pra}. As for experimental demonstrations, DQKD
protocols without entanglement, e.g., the single-photon two-state
DQKD \cite{cai04} and the single-photon four-state DQKD
\cite{deng04} (we call it the four-state protocol hereafter), were
proposed. Although the security of the four-state protocol against
some special individual attacks has been considered \cite{lm05}, its
security against general attacks has not been proved \cite{bf08}.

The difficulty of the security proof for the DQKD against general
attacks is due to the use of a two-way quantum channel. In BB84, a
qubit just travels from Alice to Bob once, carrying one-bit secret
 information. Upon receiving the qubits, Bob measures it in
either of the alternative bases to obtain Alice's key bits. A
powerful eavesdropper, Eve, whose capacity is only limited by the
physical laws, may attack the information-carrying qubit in the
one-way quantum channel. In the DQKD \cite{bf02}, however, a qubit
departs from Bob to Alice (the forward channel,  Bob-Alice), and
then it carries Alice's secret key bits back to Bob (the backward
channel, Alice-Bob). In this case, Eve might attack the qubits
traveling on both the Bob-Alice and Alice-Bob channels. Comparing it
with BB84, the security analysis of DQKD is complicated and its
unconditional security has not been proved before \cite{bf08}. In
fact, the security of two-way DQKD has been challenged over time
(see, e.g., \cite{cai03,Zhang2004,Zhang2005,Cai2006}). Some of these
challenges have led to refinement of the protocol and some have been
refuted \cite{bf08}. In this paper, we present a security proof of
the four-state protocol against general attacks, thus, confirming
the unconditional security of the protocol.

This paper is organized as follows. In Sec. \ref{Sec:twoway:4state},
we introduce the four-state protocol. Next in Sec.
\ref{Sec:twoway:security}, we present the security proof and final
key generation rate against collective attacks. Then we extend our
security proof and key generation against general attacks. We
finally conclude in Sec. \ref{Sec:twoway:conclusion}.

\section{Four-state protocol} \label{Sec:twoway:4state}
The four-state protocol works as follows.
\begin{enumerate}
\item
Bob prepares $n$ qubits randomly in one of the four states,
$|0\rangle $, $|1\rangle $, $|+\rangle $, and $|-\rangle $, where
$|\pm \rangle =(|0\rangle \pm |1\rangle)/\sqrt{2}$ and sends them to
Alice.

\item
In the check mode, Alice randomly measures part of the received
states in the $X$ or $Z$ basis.

\item
In the encoding mode, Alice randomly performs the unitary
operations $I=|0\rangle \langle0|+|1\rangle \langle 1|$ (bit 0) or
$Y=|0\rangle \langle1|-|1\rangle \langle 0|$ (bit 1) on the rest
received states.

\item
Alice sends the encoded qubits back to Bob. It is interesting to
note that $Y\left\{ |0\rangle,|1\rangle \right\} =\left\{ -|1\rangle
,|0\rangle \right\} $, and $Y\left\{ |+\rangle ,|-\rangle \right\}
=\left\{ |-\rangle ,-|+\rangle \right\} $, so Bob measures each
qubit in the same basis as the one he used for preparation. In this
way, Bob can obtain Alice's key bits deterministically, without basis
reconciliation \cite{cai04,deng04,lm05}.

\item
After Bob measures all returned qubits, Alice announces her
measurement results in the check mode. They compute the fidelity of
the forward states with results of consistent-basis measurements,
i.e., Alice measures the forward state in the same basis as Bob's
preparing it. For instance, when Bob sends a state $|0\rangle$ and
Alice measures it in the $Z$-basis in check mode, with a probability of
$f_0$, Alice's measurement outcome is $|0\rangle$. Similarly, they
can calculate the fidelity $f_1$, $f_+$, and $f_-$ of $|1\rangle$,
$|+\rangle$, and $|-\rangle$, respectively. Alice and Bob discard
results from the inconsistent-basis measurements.

\item
Alice announces partial of her key bits in the encoding mode. They
compute the error rate $e$ in the Alice-Bob channel.

\item
If Alice and Bob find the error rates in the Bob-Alice channel are
not too high (satisfying Eq. (\ref{boundaryc})), they will continue
the protocol, i.e., Alice and Bob will perform error correction (EC)
and privacy amplification (PA) to gain the secure final key bits.
Otherwise, we assume Alice and Bob will abort the protocol.

\end{enumerate}

\section{Security analysis} \label{Sec:twoway:security}

\subsection{Eve's attack in the Alice-Bob channel} Suppose that Eve only attacks
the qubits in the Alice-Bob channel. Bob prepares the forward qubits
randomly in the state $|0\rangle $, $|1\rangle $, $|+\rangle$, and
$|-\rangle $ with the same probability, i.e., the forward qubit is
prepared in a mixed state, $\rho^B=(|0\rangle\langle0|
+|1\rangle\langle1|+|+\rangle\langle+|+|-\rangle\langle-|)/4
=(|0\rangle\langle0|+|1\rangle\langle1|)/2$. To gain Alice's key
bits information, Eve has to distinguish Alice's encoded qubit
$\rho_0^B=I\rho^BI$ from $\rho_1^B=Y\rho^BY $ in the Alice-Bob
channel. Since
$\rho_0^B=\rho_1^B=(|0\rangle\langle0|+|1\rangle\langle1|)/2$, Eve
cannot gain any information about Alice's key bits if she only
attacks the qubits after Alice's encoding operation. Therefore, we
can conclude that Eve has to attack the qubits traveling on both the
Bob-Alice and Alice-Bob channels in order to gain Alice's key bits.

\subsection{Eve's attack in the Bob-Alice Channel}
\label{subsec-Eve-attack-B-to-A} Eve's most general quantum
operation can be described by a unitary operation together with an
ancilla \cite{nc}. In the Bob-Alice channel, when Bob sends a qubit
in state $|0\rangle$ and Alice measures in the basis $|0\rangle,
|1\rangle$, she will get the measurement outcomes $|0\rangle$ with
probability $c_{00}^2$, or $|1\rangle$ with probability $c_{01}^2$.
Define $f_0=c_{00}^2$ as the fidelity of state $|0\rangle$, which
can be verified by Alice and Bob in their post-processing.
Similarly, Alice and Bob can obtain $f_1=c_{11}^2$, $f_+=c_{++}^2$,
and $f_-=c_{--}^2$, respectively. Since the state space of the
forward qubit is two-dimensional, Eve's most general attack in the
Bob-Alice channel can be written in the form,
\begin{eqnarray}
\begin{array}{l}
 U_{BE}|0\rangle_{B}|E\rangle
=c_{00}|0\rangle_{B}|E_{00}\rangle+c_{01}|1\rangle_{B}|E_{01}\rangle,
\\
U_{BE}|1\rangle_{B}|E\rangle
=c_{11}|1\rangle_{B}|E_{11}\rangle+c_{10}|0\rangle_{B}|E_{10}\rangle,
\label{u01}
\end{array}
\end{eqnarray}
and
\begin{eqnarray}
\begin{array}{l}
 U_{BE}|+\rangle_{B}|E\rangle
=c_{++}|+\rangle_{B}|E_{++}\rangle+c_{+-}|-\rangle_{B}|E_{+-}\rangle,
\\
U_{BE}|-\rangle_{B}|E\rangle
=c_{--}|-\rangle_{B}|E_{--}\rangle+c_{-+}|+\rangle_{B}|E_{-+}\rangle,
\label{u+-}
\end{array}
\end{eqnarray}
where $c_{ij}$ and $c_{\pm\pm}$ can be treated as no-negative real
numbers \footnote{In general, the coefficients are complex number
and $|E_{ij}\rangle$ are normalized vectors. For simplicity, we can
rewrite the formula where the phase of each coefficient is absorbed
into the companying. For instance, $c_{ij}|E_{ij}\rangle
=\sqrt{c_{ij}c_{ij}^{*}} e^{i\delta_{ij}}|E_{ij}\rangle
=\sqrt{c_{ij}c_{ij}^{*}}(e^{i\delta_{ij}} |E_{ij}\rangle)$. Thus, we
can get that $c_{ij}\equiv \sqrt{c_{ij}c_{ij}^{*}}$ are non-negative
real number and $|E_{ij}\rangle\equiv
e^{i\delta_{ij}}|E_{ij}\rangle$ are normalized vectors.}, and
$|E_{ij}\rangle$ and $|E_{\pm\pm}\rangle$
 forms four pairs of normalized vectors. For now, we consider the
 case that Eve performs
a collective attack, i.e., $U_{BE}$ are the same for all qubits.
This restriction can be removed with quantum de Finetti theorem
\cite{qdf,renner,ckr}, and then we can prove the four-state protocol
is secure against general attacks.

As discussed above, Bob's forward qubit is prepared in a mixed state
$\rho^B=(|0\rangle\langle0|+|1\rangle\langle1|)/2$. After Eve's
attack in the Bob-Alice channel, the joint state of the forward
qubit and Eve's ancilla becomes
\begin{align*}
\rho_{Bob-Alice}^{BE} & =U_{BE}\left (\rho^B\otimes |E\rangle\langle
E|\right ) U_{BE}.
\end{align*}

After receiving the forward qubits, in the encoding mode,
Alice will encode her key bits onto the forward qubit.
With probability
$p=1/2$, she encodes key bit 0 by the operation $I_B$ or
key bit 1 by the operation $Y_B$. After the encoding,
the state of the qubit and Eve's ancilla
becomes
\begin{equation}
\rho^{ABE}=\frac{1}{2}|0\rangle\langle0|^{A}\otimes\rho_{0}^{BE}+\frac
{1}{2}|1\rangle\langle1|^{A}\otimes\rho_{1}^{BE},  \label{cqq}
\end{equation}
where $\rho^{BE}_0=\rho_{Bob-Alice}^{BE}$, and
$\rho^{BE}_1=Y_B\rho_{Bob-Alice}^{BE}Y_B$. Next, Alice sends the
encoded qubits back to Bob.

After Bob measured all the returned qubits, Alice will announce her
measurement outcomes in the check mode, so that they can gain the
fidelity of Bob's forward states, $f_{0}$, $f_{1}$, $f_{+}$, and
$f_{-}$ in the Bob-Alice channel. Alice will also publish some of
her key bits to gain the error rate $e$ of the key bits in the
Alice-Bob channel. For simplicity, we will first consider the case
that $f_0=f_1$ and $f_+=f_-$.

\subsection{Secret key generation}
The asymptotic key generation rate can be defined as $r=\lim_{m\rightarrow
\infty }{k(m)}/{m}$, where $m$ is the size of the raw key and
$k(m)$ is the number of the final key bits. Alice sends Bob EC
information over a classical channel so that he can correct his raw
key to match Alice's.  This EC information is encrypted using
pre-shared secret key bits and thus is unknown to Eve.  The final
key is then derived by applying two-universal hashing to their
common raw key as PA \cite{rk05}. In the asymptotic scenario,
the secure key rate $r_{PA}$ for secret key generation is bounded by the conditional
entropy of Alice and Bob's key bits given the quantum information of
Eve about the key bits, $r_{PA}=S(\rho^A|\rho^{BE})$.

After Alice's encoding operations, Eve can gain some quantum
information about Alice's key bit from the quantum state
$\rho^{BE}=tr_A{\rho^{ABE}}$ that is a joint state of the backward
qubit and her ancilla. Here, we assume the worst case that Eve uses
the entire state $\rho_{BE}$ to gain information about the key bit,
even though she may have to send part of the state to Bob. With
Renner and K\"onig's results \cite{rk05}, we have
$r_{PA}=S(\rho^A|\rho^{BE})=S(\rho^{ABE})-S(\rho^{BE})$, where
$S(\rho^{ABE})=-tr\rho^{ABE}\log_2\rho^{ABE}$, and
$S(\rho^{BE})=-tr\rho^{BE}\log_2\rho^{BE}$. In the following, we
should calculate the eigenvalues of $\rho^{ABE}$ and $\rho^{BE}$ to
get $S(\rho^{ABE})$ and $S(\rho^{BE})$.

\subsection{Key generation rate for PA}

Let us denote that
\begin{equation}\label{twoway:security:Estate}
\begin{aligned}
\langle E_{00}|E_{01}\rangle &= s_0+is_1, \\
\langle E_{00}|E_{10}\rangle&=u_0+iu_1, \\
\langle E_{00}|E_{11}\rangle&=p_0+ip_1, \\
\langle E_{11}|E_{10}\rangle&=r_0+ir_1, \\
\langle E_{01}|E_{11}\rangle&=v_0+iv_1, \\
\langle E_{01}|E_{10}\rangle&=q_0+iq_1, \\
\end{aligned}
\end{equation}
where $p_i$, $q_i$, $r_i$, $s_i$, $u_i$ and $v_i$ are real number.
Taking the inner product of two equations of Eq.(\ref{u01}) gives
\begin{equation}
c_{00}c_{10}\langle E_{10}|E_{00}\rangle+c_{11}c_{01}\langle
E_{01}|E_{11}\rangle=0. \label{c01}
\end{equation}
For simplicity, we rewrite $c_0\equiv c_{00}=c_{11}$ and $c_1 \equiv
c_{01}=c_{10}$ and thus we have $u_0=-v_0$ and $u_1=-v_1$.

Let us calculate $S(\rho^{ABE})$ and $S(\rho^{BE})$. A
straightforward computation \footnote{First, we select a completely
orthogonal basis $|E_{00}'\rangle$, $|E_{01}'\rangle$,
$|E_{11}'\rangle$, and $|E_{10}'\rangle$ on the Hilbert space
$\mathscr{H}^E$. Then we have
$|E_{00}\rangle=\sum_{ij}a_{ij}|E_{ij}'\rangle$,
$|E_{01}\rangle=\sum_{ij}b_{ij}|E_{ij}'\rangle$,
$|E_{11}\rangle=\sum_{ij}f_{ij}|E_{ij}'\rangle$ and
$|E_{10}\rangle=\sum_{ij}g_{ij}|E_{ij}'\rangle$, where $i,j\in
\{0,1\}$, and $\langle E_{00}|E_{01}\rangle=s_0+is_1$, $\langle
E_{00}|E_{10}\rangle=-\langle E_{01}|E_{11}\rangle=u_0+iu_1$,
$\langle E_{00}|E_{11}\rangle=p_0+ip_1$, $\langle
E_{11}|E_{10}\rangle=r_0+ir_1$ and $\langle
E_{01}|E_{10}\rangle=q_0+iq_1$. After some tedious calculations, we
can gain the eigenvalues of $\rho^{ABE}$ on the basis $|0\rangle_A$,
$|1\rangle_A$, $|0\rangle_B$, $|1\rangle_B$, $|E_{00}'\rangle$,
$|E_{01}'\rangle$, $|E_{11}'\rangle$, and $|E_{10}'\rangle$ in the
Hilbert space
$\mathscr{H}^{ABE}=\mathscr{H}^A\otimes\mathscr{H}^B\otimes\mathscr{H}^E$,
and the eigenvalues of $\rho^{BE}$ on the basis $|0\rangle_B$,
$|1\rangle_B$, $|E_{00}'\rangle$, $|E_{01}'\rangle$,
$|E_{11}'\rangle$, and $|E_{10}'\rangle$ in the Hilbert space
$\mathscr{H}^{BE}=\mathscr{H}^B\otimes\mathscr{H}^E$.} shows that
the eigenvalues of $\rho^{ABE}$ can be obtained as
$\lambda_{0,1,2,3,4,5,6,7,8,9,10,11}^{ABE}=0$ and
$\lambda_{12,13,14,15}^{ABE}={1}/{4}$. Thus, we obtain
$S(\rho^{ABE})=\sum_{i}\lambda_{i}^{ABE}\log_{2}\lambda_{i}^{ABE}=2$,
where we have used the convention $0\log_{2}0=0$. The eigenvalues of
$\rho^{BE}=tr_{A}\rho^{ABE}$ are $\lambda_{0,1,2,3,}^{BE}=0$,
$\lambda_{4}^{BE}=\frac{1}{4} [1+(\Delta_{1}+\Delta_2)]$,
$\lambda_{5}^{BE}=\frac{1}{4} [1+(\Delta_{1}-\Delta_2)]$,
$\lambda_{6}^{BE}=\frac{1}{4} [1-(\Delta_{1}+\Delta_2)]$ and
$\lambda_{7}^{BE}=\frac{1}{4} [1-(\Delta_{1}-\Delta_2)]$, where
$\Delta_1=\sqrt{(c_0^2p_0+c_1^2q_0)^2+(c_{0}^2p_1+c_1^2q_1)^2
+c_0^2c_1^2(s_0+r_0)^2}$ and
$\Delta_2=\sqrt{c_0^2c_1^2(s_0-r_0)^2}\,$ \footnote{Calculation of
the eigenvalues of $\rho^{ABE}$ and $\rho^{BE}$ was partially
performed by computers with symbolic computation. It is surprising
at first glance that the parameters $u_{i}$, $v_{i}$, $s_{1}$ and
$r_{1}$ are canceled out in the formula for the eigenvalues of
$\rho^{BE}$. The reason that $u_{i}$ and $v_{i}$ are canceled out is
that $u_{i}$ and $v_{i}$ are symmetric with $u_{i}=-v_{i}$. Also,
$s_{1}$ and $r_{1}$ are canceled out because $s_{1}$ and $r_{1}$ are
the imaginary parts of $\langle E_{00}|E_{01}\rangle$ and $\langle
E_{11}|E_{10}\rangle$, respectively, and thus they can be canceled
out by diagonalization as arbitrary phase factors. Similarly, the
eigenvalues of $\rho^{ABE}$ are independent of the parameters
$u_{i}$, $v_{i}$, $r_{i}$, $s_{i}$, $p_{i}$ and $q_{i}$.}.
Considering the concavity of von Neumann entropy \cite{nc}, we can
find $S(\rho^{BE})=-\sum_{i}
\lambda_{i}^{BE}\log_{2}\lambda_{i}^{BE}$ approaches its maximum
when $r_0=s_0=q_1=p_1=0$. In this case, we have
$\lambda_{4,5}^{BE}=\frac{1}{4}[1+(c_0^2p_0+c_1^2q_0)]$ and
$\lambda_{6,7}^{BE}=\frac{1}{4}[1-(c_0^2p_0+c_1^2q_0)]$ \footnote{As
discussed earlier, Eve has to attack the traveling qubit twice, on
both Bob-Alice and line Alice-Bob channels, to gain Alice's key
bits. On the Bob-Alice channel, Eve attacks the forward qubit to
distinguish its states, and then she can determine Alice's encoding
operations after she attacked the backward qubit on Alice-Bob
channel. It is optimal for Eve to distinguish Alice's states when
$\langle E_{00}|E_{01}\rangle=\langle E_{00}|E_{10}\rangle=\langle
E_{11}|E_{10} \rangle=\langle E_{01}|E_{11}\rangle=0$, $\langle
E_{00}|E_{11}\rangle=p_0$, and $\langle E_{01}|E_{10}\rangle=q_0$
\cite{grtz}, and thus it is optimal among collective attacks in the
four-state protocol.}.

With the eigenvalues of $\rho^{ABE}$ and $\rho^{BE}$, we get
\begin{align*}
&S(\rho^{ABE})=2,\\
&S(\rho^{BE})=-\sum_i\lambda_i^{BE}\log_{2}\lambda_i^{BE}.
\end{align*}
From Eqs. (\ref{u01}) and (\ref{u+-}), we can get
\begin{eqnarray}
2c_{++}|E_{++}\rangle & =c_{00}|E_{00}\rangle +c_{01}|E_{01}\rangle
+c_{11}|E_{11}\rangle +c_{10}|E_{10}\rangle. \notag \\
2c_{+-}|E_{+-}\rangle & =c_{00}|E_{00}\rangle -c_{01}|E_{01}\rangle
-c_{11}|E_{11}\rangle +c_{10}|E_{10}\rangle, \notag \\
2c_{--}|E_{--}\rangle & =c_{00}|E_{00}\rangle -c_{01}|E_{01}\rangle
+c_{11}|E_{11}\rangle -c_{10}|E_{10}\rangle, \notag \\
2c_{-+}|E_{-+}\rangle & =c_{00}|E_{00}\rangle +c_{01}|E_{01}\rangle
-c_{11}|E_{11}\rangle -c_{10}|E_{10}\rangle. \notag
\end{eqnarray}
When considering $f_+=f_-$, the equations above give a
\textit{crucial} boundary condition,
\begin{equation}
1+c_{0}^2p_0+c_{1}^2q_0=2c_{++}^2.  \label{bc}
\end{equation}
Let us analyze the maximum of $S(\rho^{BE})$ when the fidelity,
$c_{0}^2$ and $c_{++}^2$, were verified by Alice and Bob in their
post-processing.

With the boundary condition of Eq.(\ref{bc}) and $-1\leq p_{0},
q_{0}\leq 1$, after some tedious calculation, we obtain that
$c_0^2p_0+c_1^2p_1 \geq 2c_{++}^2-1-2c_{1}^2$ \footnote{With the
boundary condition $1+c_{0}^2p_0+c_{1}^2q_0=2c_{++}^2$, we can get
$2c_{++}^2-1-c_{0}^2p_0=c_{1}^2q_0\leq c_{1}^2|q_0|\leq c_{1}^2$,
i.e., $c_{0}^2p_0\geq 2c_{++}^2-1-c_{1}^2$, since $|p_0|\leq 1$ and
$|q_0|\leq 1$. Thus, we can find that $c_{0}^2p_0+c_{1}^2q_0\geq
c_{0}^2p_0-c_{1}^2|q_0|\geq c_{0}^2p_0-c_{1}^2\geq
2c_{++}^2-1-2c_{1}^2$. Here $2c_{++}^2-1-2c_{1}^2\geq 0$, i.e.,
$c_{++}^2-c_{1}^2\geq 1/2$, is required and should be verified by
Alice and Bob in their post-processing.} and $S(\rho^{BE})$
approaches the maximum when $c_0^2p_0+c_1^2p_1
=2c_{++}^2-1-2c_{1}^2\geq 0$. After verifying the condition
$c_{++}^2-c_{1}^2\geq 1/2$ in their post-processing, Alice and Bob
can obtain
\begin{eqnarray}
S(\rho^{BE}) &=&\max \left\{
S(\rho^{BE}|p_{0},q_{0})\right\} \notag \\
=&-&\frac{2c_{++}^{2}-2c_{1}^2}{2}\log _{2}\frac{
2c_{++}^{2}-2c_{1}^2}{4}  \notag \\
&-&\frac{2-2c_{++}^{2}+2c_{1}^2}{2}\log _{2}\frac{
2-2c_{++}^{2}+2c_{1}^2}{4}. \label{ssy}
\end{eqnarray}
Therefore, after verifying $c_{++}^2-c_{1}^2\geq 1/2$ Alice and Bob
can get the rate of PA against collective attacks,
\begin{eqnarray}
r_{PA}(\xi)=S(\rho^A|\rho^{BE})=1-h(\xi), \label{rsy}
\end{eqnarray}
where $\xi=c_{++}^2-c_{1}^2$, and
$h(x)=-x\log_{2}x-(1-x)\log_{2}(1-x)$ is the binary Shannon entropy.

In particular, if Eve does not attack the forward qubits in the
Bob-Alice channel, i.e., $f_{0}=f_{1}=f_{+}=f_{-}=1$, one can find
that $r_{PA}(\xi)=1$. This states that Eve cannot gain any
information about Alice's key bits if she doesn't attack the travel
qubit in the Bob-Alice channel first.

Consider the case that Eve measures each forward qubit in the
Bob-Alice channel in the basis $|0\rangle, |1\rangle$. Alice and Bob
can verify that $f_{0}=f_{1}=1$, and $f_{+}=f_{-}=1/2$. In this
case, we have $r_{PA}(\xi)=0$. On the other hand, Eve can also
measure each forward qubit in the Bob-Alice channel in the basis
$|+\rangle, |-\rangle$, which gives $f_{+}=f_{-}=1$ and
$f_{0}=f_{1}=1/2$, and thus $r_{PA}(\xi)=0$. That is, Eve can gain
full information of Alice's key bits if she has exactly known the
forward states before Alice's encoding operations.

\subsection{PA for the practical quantum channels}

For the practical quantum channels, the condition $f_0=f_1$ and
$f_+=f_-$ is too strict to be satisfied. We can use the following
strategy to symmetrize Eve's channels and eliminate this condition.
We first identify four locations --- Alice's side and Bob's side of
the Bob-Alice and Alice-Bob channels. For each bit, we randomly
insert a bit flip operation $Y$ at these four locations. So the four
locations are either all $Y$ or all $I$. In this way, the fidelities
of the new the Bob-Alice channel are simply the average of that of
the original line [i.e., the new $f_0$ and $f_1$ ($f_+$ and $f_-$)
are the average of the old $f_0$ and $f_1$ ($f_+$ and $f_-$)]~
\footnote{ Essentially, Bob-Alice channel in Eqs.~\eqref{u01} and
\eqref{u+-} is symmetrized by
 $U_{BE}|\psi\rangle_{B}|E\rangle |0\rangle_{B'}
 + Y_B U_{BE} Y_B |\psi\rangle_{B}|E\rangle |1\rangle_{B'}$
 where system $B'$ indicates whether we have a bit flip or not.
 We can assume that we give $B'$ to Eve after the qubit transmission
 so that the pure-state analysis in the previous sections still applies.
 Note that giving $B'$ to Eve afterwards means that $U_{BE}$ and system
 $E$ are independent of our extra bit flip.}.
Thus, this justifies the conditions $f_0=f_1$ and $f_+=f_-$ used in
the previous sections.

Now, let us simplify further. Consider the two Y operations at
Alice's side located before and after Alice's encoding operation.
Since Alice's encoding performs either $Y$ or $I$, the two new $Y$
operations commute with Alice's encoding operation and cancel out.
Now consider the two $Y$ operations at Bob's side, if he originally
wants to send bit $b=0,1$ in basis $W=X,Z$ through the the Bob-Alice
channel, he now sends bit $1-b$ in basis $W$ to implement $Y$. When
he receives the qubit from the channel Alice-Bob, he performs $Y$ on
the incoming qubit, measures in basis $W$, records the bit $b'$, and
computes $b\oplus b'$ as Alice's key bit. This is the same as his
measuring the incoming qubit in $W$, flipping the bit result $1-b'$
to $b'$, and computing $b\oplus b'$ as Alice's key bit. Note that
the input to the raw Bob-Alcie channel is $1-b$ in the basis $W$ and
the output of the raw Alice-Bob channel is measured in $W$ resulting
in bit $1-b'$. Thus, Bob can use this information directly to infer
Alice's key bit as $(1-b)\oplus (1-b')=b\oplus b'$ without actually
implementing the two $Y$ operations. Since $b$ is uniformly
distributed, so is $1-b$ and the original protocol is recovered
(i.e., no need to introduce the extra four $Y$ operations) except
that the averages of the fidelities (instead of the individual
fidelities) determined in the check mode are used in the key
generation formulas. Specifically, we can use
$\xi=f_{+,-}+f_{0,1}-1$ in Eq.~(\ref{rsy}) to calculate PA for any
quantum channels in the four-state protocol, where
$f_{0,1}=(c_{00}^2+c_{11}^2)/2$, $f_{+,-}=(c_{++}^2+c_{--}^2)/2$,
and $c_{ij}^2$'s are the original fidelities of the Bob-Alice
channel determined in the check mode.

\subsection{Final key generation rate}

In the post-processing, Alice and Bob should estimate the fidelity
in the Bob-Alice channel, and $e$ in the Alice-Bob channel. Then
they will perform EC and PA to generate the final key bits. In an
asymptotic scenario, after verifying
\begin{equation} \label{boundaryc}
f_{+,-}+f_{0,1}\geq 3/2,
\end{equation}
Alice and Bob can obtain the secure final key against collective
attacks with the generation rate
\begin{equation}
r=1-h(\xi)-h(e), \label{fk}
\end{equation}
where $h(e)$ is the amount of key bits Alice and Bob should
sacrifice in EC. To compare the key rate performance of the two-way
DQKD given by Eq.~\eqref{fk} to that of BB84 given by $1-2h(e)$, we
can assume a symmetric attack by Eve in the DQKD case (in which
$c_{++}^2=1-e$ and $c_1^2=e$) so that $\xi=1-2e$. Thus, the key rate
of DQKD is $1-h(2e)-h(e)$ which is smaller than that of BB84 when
there are errors.

\subsection{Security against general attacks} Our analysis above
assumes that the state $\rho^{ABE}$ of each run of the system is
independent of and identical to the states of other runs, i.e., we
assume the entire state for the $n$ runs is $(\rho^{ABE})^{\otimes
n}$. This collective-attack result can be extended to general
attacks where the entire state is arbitrary without any restriction
on the $n$ subsystems. Note that the four-state protocol is
unchanged with a randomized permutation step. In light of
ref.~\cite{renner,ckr}, after permuting the $n$ subsystems and
discarding $k$ of them, the resulting $n-k$ subsystems can be
approximated by $n-k$ independent and identically distributed
subsystems \footnote{The quantum de Finetti theorem applies only to
finite dimensional Hilbert spaces. In our proof, Eve's state is a
purification of Alice and Bob's state. Since the latter is finite
dimensional, the former is also finite dimensional. Thus, Alice and
Bob's knowledge about Eve's state is finite dimensional, despite
that Eve's knowledge about her own state could still be infinite
dimensional. This can be seen by the finite number of eigenvalues of
the joint ABE state given in the paragraph below Eq.~\eqref{c01}.
Also, the Stinespring dilation theorem states that Eve's effective
ancilla is no more than four-dimensional since the dimensional of
the Hilbert spaces of travel qubit  is two (see
Sec.~\ref{subsec-Eve-attack-B-to-A}). Therefore, the application of
the quantum de Finetti theorem is valid in our proof.}. Therefore,
the final key bits in our proof with the generation rate in
Eq.(\ref{fk}) are secure against general attacks.

\section{Conclusion} \label{Sec:twoway:conclusion}

We have proved that the four-state protocol is secure against
general attacks, thus ending the long-standing dispute about the
security of the deterministic QKD protocols. Our work may be
extended to other QKD protocols with a two-way quantum channel and
shine new light on the universality of QKD.

\section{Acknowledgement}

We are grateful to Marco Lucamarini and Baocheng Zhang for the
helpful discussions. QYC especially thanks Hoi-Kwong Lo for his
hospitality when visiting University of Toronto. Financial support
from RGC Grants No.~HKU 701007P and 700709P of the HKSAR government
and NSFC Grant No. 11074283, is gratefully acknowledged.

\end{document}